\documentclass[aps,prl,reprint,superscriptaddress,longbibliography,nofootinbib]{revtex4-2}
\usepackage[T1]{fontenc}
\usepackage{newtxtext,newtxmath}
\usepackage{amsmath,graphicx,xcolor,url}
\usepackage[hidelinks]{hyperref}
\graphicspath{{}}
\newcommand{\pnl}[1]{\textbf{#1}}
\begin{document}
	
	\title{Mode locking by phase-sensitive amplification}
	
	\author{Francesco De Lucia}
	\thanks{F.D.L. and Y.S. contributed equally to this work.}
    \affiliation{Service OPERA-Photonique, Universit\'e Libre de Bruxelles, B-1050 Brussels, Belgium}
	\affiliation{University of Lille, CNRS, UMR 8523 PhLAM, F-59000 Lille, France}
	\author{Yifan Sun}
    \email{yifan.sun@ulb.be}	
	\affiliation{Service OPERA-Photonique, Universit\'e Libre de Bruxelles, B-1050 Brussels, Belgium}
	\author{Georges Semaan}
	\affiliation{Service OPERA-Photonique, Universit\'e Libre de Bruxelles, B-1050 Brussels, Belgium}
	\author{Clement Dupont}
	\affiliation{Service OPERA-Photonique, Universit\'e Libre de Bruxelles, B-1050 Brussels, Belgium}
	\author{Nicolas Englebert}
	\affiliation{Department of Electrical Engineering, California Institute of Technology, Pasadena, California 91125, USA}
	\author{Simon-Pierre Gorza}
	\affiliation{Service OPERA-Photonique, Universit\'e Libre de Bruxelles, B-1050 Brussels, Belgium}
	\author{Fran\c{c}ois Leo}
	\email{francois.leo@ulb.be}
	\affiliation{Service OPERA-Photonique, Universit\'e Libre de Bruxelles, B-1050 Brussels, Belgium}
	
	\begin{abstract}
        Mode locking is the mechanism by which lasers organize many optical modes into stable trains of ultrashort pulses\cite{haus2000mode,Keller2003,Fermann2013,sun2019mode}. It underpins optical frequency combs, precision timing and spectroscopy\cite{udem2002optical,Diddams2020}, and provides the phase-coherent waveforms required for applications ranging from strong-field physics to optical frequency metrology\cite{Baltuska2003,Fortier2011}.
		Conventional pulse-forming mechanisms select pulse amplitude without fixing its phase relative to an external optical reference, leaving pulse formation and optical-phase stabilization as separate functions~\cite{Telle1999,Jones2000,Fortier2019}.
		Phase-sensitive amplification offers a different possibility: the optical reference can participate directly in pulse formation~\cite{Longhi1995,Kutz2008}.
		Here we demonstrate this principle in a continuously pumped oscillator.
		Kerr-induced phase evolution and cavity detuning turn pump-referenced parametric gain into a power-dependent discriminator that supports stable single-soliton operation.
		Roundtrip-resolved interferometry reveals that the soliton optical phase remains referenced to the master laser modulo $\pi$.
		These results establish phase-sensitive mode locking, in which pulse-amplitude selection and optical-phase referencing arise from the same coherent nonlinear feedback.
	\end{abstract}
	
	\maketitle

\begin{figure*}[!htbp]
		\centering
		\includegraphics[page=1,width=\linewidth]{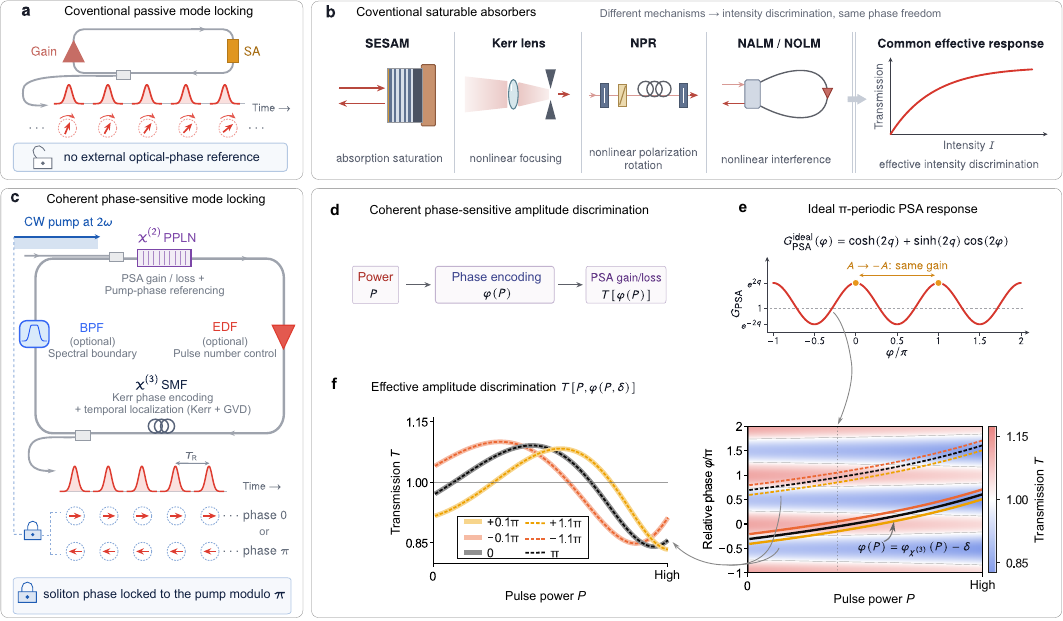}
		\caption{\textbf{From conventional to phase-sensitive mode locking.}
			\pnl{a}, Conventional passive mode locking selects pulses through intensity-dependent loss while leaving their optical phase free relative to an external reference.
			\pnl{b}, Semiconductor saturable absorber mirrors (SESAM), Kerr lenses, nonlinear polarization rotation (NPR), and nonlinear amplifying or optical loop mirrors (NALM/NOLM) realize intensity discrimination through different mechanisms; the common response is schematic.
			\pnl{c}, In the continuously pumped fiber--PPLN oscillator, $\chi^{(2)}$ phase-sensitive amplification supplies gain and an optical phase reference. Kerr nonlinearity and anomalous dispersion in single-mode fiber (SMF) support pulse localization. Auxiliary erbium-doped fiber (EDF) gain, operated below its independent lasing threshold, partially compensates cavity loss and couples the pulse population; the bandpass filter (BPF) suppresses higher-order parametric resonances.
			\pnl{d}, Coherent amplitude discrimination: Kerr propagation encodes pulse power into phase, cavity detuning sets a bias, and phase-sensitive amplification (PSA) converts the resulting pump-referenced phase into gain or loss.
			\pnl{e}, Ideal undepleted PSA gain is $\pi$ periodic, giving equal gain to the two sign-related states $A$ and $-A$; at fixed phase its fractional gain is amplitude independent.
			\pnl{f}, Illustrative effective transmission (left) along power-dependent phase trajectories through the phase-sensitive gain landscape (right). Suitable biases give weak fields net attenuation, intermediate powers amplification and a high-power rollover; other biases, such as $-0.1\pi$, amplify weak fields and do not provide this discrimination window. Dashed trajectories are the $\pi$-shifted counterparts of solid trajectories and have identical transmission. The unity line separates net gain from loss. These curves illustrate single-pass discrimination and are not calibrated full-cavity energy responses.
			CW, continuous wave; GVD, group-velocity dispersion; PPLN, periodically poled lithium niobate.}
		\label{fig:concept}
	\end{figure*}

	
	Mode locking organizes optical modes into ultrashort pulses, underpinning frequency combs, precision timing and spectroscopy~\cite{haus2000mode,Keller2003,udem2002optical,Diddams2020}. Its nonlinear pulse-forming mechanisms select pulse amplitude and establish the relative phases among optical modes needed to form a localized waveform, but generally leave the pulse's common optical phase free to evolve relative to an external optical reference. 
    Conventional passive discriminators, including saturable absorbers, Kerr lenses, nonlinear polarization rotation and loop mirrors, are insensitive to this common phase (Fig.~\ref{fig:concept}a,b)~\cite{Grelu2012,Keller1996,Spence1991,Hofer1991,Doran1988,Fermann1990}. Establishing a stable optical-phase relationship, essential for coherent optical-frequency synthesis, therefore ordinarily requires separate measurement and feedback~\cite{Telle1999,Jones2000,Fortier2019}.
	
	Coherently driven resonators bring an optical reference directly into the pulse formation dynamics. 
	Continuous-wave injection sustains solitons in passive Kerr resonators~\cite{Leo2010,Herr2014} and gain-assisted fiber, microsphere and semiconductor resonators~\cite{EnglebertActive2021,semaanCavitySolitonsUltrahigheffectiveQ2026,Liang2026,Kazakov2025}, and can synchronize soliton combs to an external optical reference~\cite{Moille2023}.
	Yet additive driving comes at a cost: the optical reference also sustains a continuous-wave background. In long fiber resonators, this background can carry most of the circulating energy despite being orders of magnitude weaker than the soliton peak~\cite{EnglebertContrast2023}.
	
	Degenerate parametric driving offers another possibility. Instead of adding a field at the signal frequency, it couples the signal to its pump-generated phase conjugate. 
    The resulting gain depends on the signal phase relative to the pump, but is identical for phases separated by $\pi$ (Fig.~\ref{fig:concept}c)~\cite{Caves1982,Wong2010,Marandi2012Coherence}. The optical reference can thus be supplied without requiring an extended coherent background. Parametrically driven Kerr cavity solitons (PDCSs) have been demonstrated in fiber and integrated resonators~\cite{Englebert2021,Moille2024}, with related developments including predicted soliton crystals~\cite{sunParametricallyDrivenKerr2026a}, topological soliton combs~\cite{Englebert2026} and self-aligned parametric microcombs~\cite{Moille2026}. Continuous-wave-pumped optical parametric oscillators can also generate mode-locked quadratic solitons~\cite{Nie2020OL,Nie2020PRApplied,Musgrave2026}. Earlier theoretical work explored how parametric gain could participate in pulse formation, from solitary-pulse generation in optical parametric oscillators~\cite{Longhi1995} to Kerr-assisted PSA as a mode-locking discriminator~\cite{Kutz2008}.
	
	Here we experimentally realize Kerr--PSA mode locking in a continuously pumped hybrid optical parametric oscillator (Fig.~\ref{fig:concept}c,d). Pulses self-start during laser–cavity frequency scans without an injected pulse seed or synchronous pumping, and can evolve into an isolated soliton. Kerr phase evolution and cavity detuning turn pump-referenced parametric gain into a power-dependent discriminator, while shared erbium-gain saturation helps select the surviving pulse population.
    Our modeling predicts locally restoring pulse-energy responses, while roundtrip-resolved interferometry reveals master-referenced optical phases in both the $\pi$- and $2\pi$-resonance classes. 
    Together, these results establish a new form of coherent mode locking in which pulse formation and optical-phase referencing arise from the same nonlinear feedback. Beyond conventional ultrafast pulse generation, this principle offers a route towards directly optically referenced and frequency-agile ultrafast sources, with potential applications in coherent frequency synthesis, precision metrology and spectroscopy.

	\section*{Coherent phase-sensitive power discrimination}
	
	Our oscillator combines a periodically poled lithium niobate (PPLN) crystal pumped at $\omega_p=2\omega_0$ with predominantly anomalous-dispersion fiber (Fig.~\ref{fig:concept}c,d). In the ideal undepleted-pump limit, the phase-sensitive input--output relation gives the pulse-energy gain $G=E_{\mathrm{out}}/E_{\mathrm{in}}=\cosh(2q)+\rho\sinh(2q)\cos(2\varphi)$ (Methods; Supplementary Note~2). Here $q$ is the parametric coupling strength, $\varphi$ is the signal phase relative to the pump, and $\rho$ measures the phase uniformity across the pulse, reaching unity for a phase-uniform pulse. The gain repeats for phase shifts of $\pi$, but is independent of signal amplitude at a fixed phase and pulse shape (Fig.~\ref{fig:concept}e). PSA is therefore phase selective, but amplitude blind.
	
    Kerr propagation supplies this missing amplitude dependence. Self-phase modulation makes the accumulated optical phase depend on pulse power (Fig.~\ref{fig:concept}d), while the balance between Kerr nonlinearity and anomalous dispersion supports a nearly chirp-free soliton, preserving the phase uniformity needed for high-contrast PSA  (Supplementary Note~2). Cavity detuning provides a tunable phase bias, shifting the range of pulse power that experiences the strongest PSA gain. At a suitable detuning, pulses within this range are amplified more strongly than weaker or stronger fields, producing a saturable-absorber-like response with a finite-amplitude discrimination window (Fig.~\ref{fig:concept}f).

	\begin{figure*}[!htbp]
		\centering
		\includegraphics[width=\linewidth]{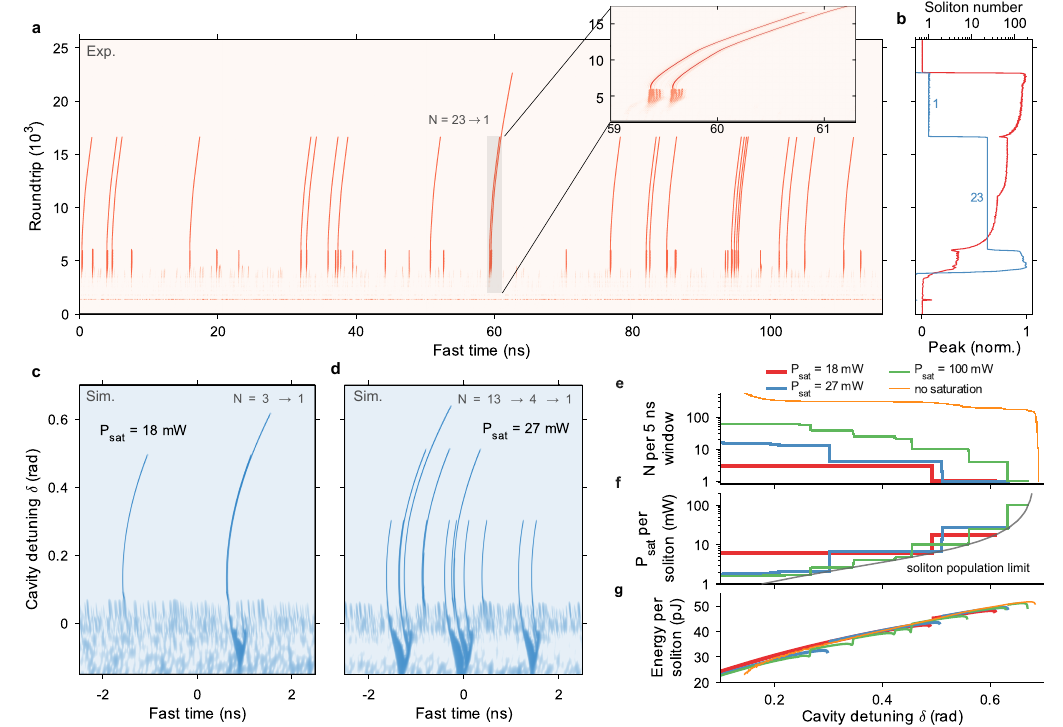}
		\caption{\textbf{Erbium gain saturation controls pulse number.}
			\pnl{a}, Measured pulse evolution during a fast forward detuning scan, resolved over consecutive roundtrips and the 116-ns cavity period. Inset, contraction of a dense pulse cluster. 
			\pnl{b}, Normalized peak signal (red, lower axis) and number of resolved pulses (blue, logarithmic upper axis), showing a 23-to-one transition. 
			\pnl{c},\pnl{d}, Numerical detuning scans with erbium-gain saturation powers $P_{\rm sat}=18$ and 27~mW, showing population sequences of three to one and 13 to four to one, respectively, within a 5-ns simulation window.
			\pnl{e}, Simulated pulse number versus detuning for $P_{\rm sat}=18$, 27 and 100~mW, and without erbium-gain saturation.
			\pnl{f}, Effective single-pulse saturation power $P_{\rm sat}/N$, where $N$ is the pulse number. Black, independently calculated single-soliton loss boundary $\delta_1(P)$; population steps follow this boundary, apart from some closely bound pulses.
			\pnl{g}, Median energy per soliton versus detuning for the same runs, showing weak dependence on erbium saturation.}
		\label{fig:selection}
	\end{figure*}    

    In contrast to nonlinear polarization rotation and nonlinear loop mirrors, where interfering fields originate from the same pulse and their common optical phase cancels from the transmission~\cite{Hofer1991,Doran1988,Fermann1990}, Kerr--PSA discrimination is referenced to an external coherent pump. The signal instead couples to its pump-generated conjugate: the gain itself depends on the optical phase relative to an external reference (Fig.~\ref{fig:concept}b,d). This coherent interaction both supplies the gain sustaining a PDCS and participates in its amplitude selection. Although pump depletion can also reduce gain at higher powers~\cite{MasArabi2023}, single-pass controls show that depletion alone cannot reproduce the rising branch of the Kerr--PSA response (Supplementary Fig.~S5).

	\section*{Fundamental phase-sensitive mode locking}
	
	Continuous-wave parametric pumping imposes no pulse-arrival slots, leaving the number of circulating solitons to be determined by the cavity dynamics. During a fast detuning scan (Fig.~\ref{fig:selection}a,b), transient clusters give way to 23 resolved peaks separated by at least 0.2~ns within the 116-ns cavity roundtrip. After approximately 16,600 roundtrips, the population collapses from 23 pulses to one, establishing fundamental-rate mode locking with a single pulse per roundtrip.  The near-simultaneous disappearance of pulses separated by much more than the approximately 15-ps walk-off-mediated pump depletion range suggests competition through a cavity-wide gain reservoir.
	
	The erbium-doped fiber provides such coupling through shared, phase-insensitive gain. Operating below its independent lasing threshold, it partially compensates cavity loss, while gain saturation couples widely separated solitons through their total circulating power~\cite{EnglebertActive2021,semaanCavitySolitonsUltrahigheffectiveQ2026,Liang2026}. Erbium gain is not required for Kerr--PSA mode locking: without saturation, its constant gain simply reduces the net cavity loss, which could instead be achieved in a sufficiently low-loss cavity. 
    In the present oscillator, shared erbium-gain saturation~\cite{Tang2005,Grelu2012} provides an additional global mechanism for selecting soliton number, complementing the local coherent discrimination produced by Kerr–PSA feedback.
	
	\begin{figure*}[th]
		\centering
		\includegraphics[width=\linewidth]{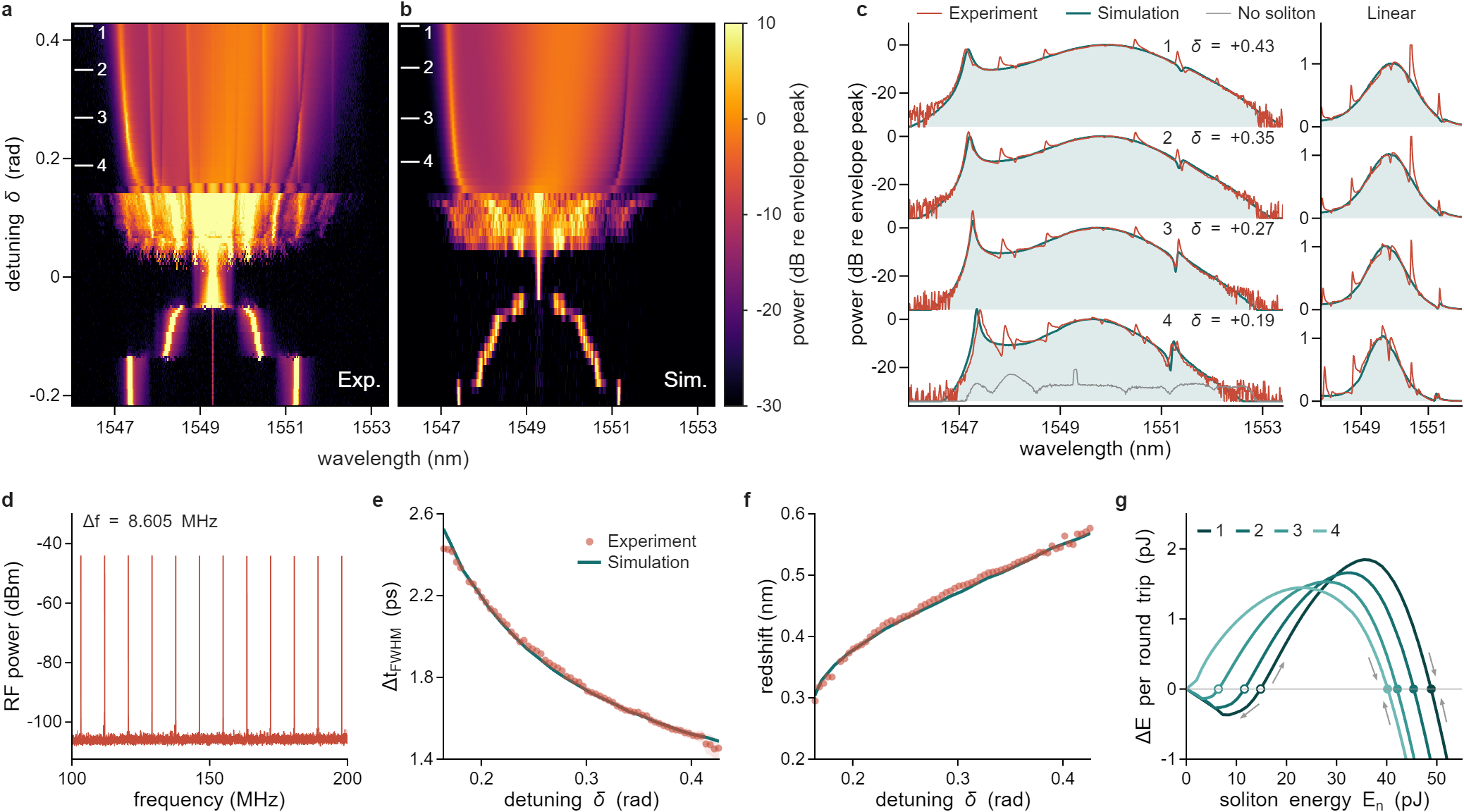}
		\caption{\textbf{A single soliton and its local energy response.}
			\pnl{a}, Measured backward detuning scan initialized from a single soliton, showing spectral broadening, a blue dispersive wave and its red parametric idler.
			\pnl{b}, Corresponding simulation from a mean-field cavity model, including the analyser response. Maps are normalized to their brightest spectral-envelope peaks; the experimental detuning origin is fitted (Methods).
			\pnl{c}, Measured (red) and simulated (teal) spectra at $\delta=0.19$, 0.27, 0.35 and 0.43~rad, normalized to their envelope peaks; right, the same spectra on linear axes. Grey, an independently measured no-soliton spectrum showing stationary spectral features.
			\pnl{d}, Radiofrequency harmonics spaced by the 8.605-MHz cavity repetition rate, from a separate acquisition.
			\pnl{e}, Transform-limited sech$^2$-equivalent duration and \pnl{f}, spectral redshift relative to 1,549.284~nm, inferred from model-assisted fits to measured spectral envelopes. Shading represents half the spread among three spectral weightings.
			\pnl{g}, Calculated one-roundtrip change in gated pulse energy after soliton-like perturbations preserving $P_{\rm peak}T_0^2$. Filled circles mark the stationary energies $E_*$. Positive values indicate net energy gain.}
		\label{fig:soliton}
	\end{figure*}
	
	We examine this role through numerical detuning scans with varying erbium saturation power $P_{\rm sat}$. Representative trajectories show populations decreasing from three to one soliton at $P_{\rm sat}=18$~mW, and from thirteen to four and then one at 27~mW (Fig.~\ref{fig:selection}c,d). Across four conditions, weaker saturation (larger $P_{\rm sat}$) permits larger populations and narrows the single-soliton detuning interval. Without gain saturation, many pulses persist until the soliton branch is lost (Fig.~\ref{fig:selection}e).
	
	The population steps follow a simple shared-gain constraint. For $N$ comparable, well-separated solitons, their combined saturation is approximately equivalent to a single soliton with effective saturation power $P_{\rm sat}/N$. Pulse losses follow the independently calculated single-soliton loss boundary $\delta_1(P_{\rm sat}/N)$ (Fig.~\ref{fig:selection}f): each loss relaxes gain saturation, allowing the survivors to persist at larger detunings. Despite the markedly different populations, the median energy per soliton follows nearly the same detuning-dependent branch (Fig.~\ref{fig:selection}g). Thus, shared gain controls how many solitons survive, with little influence on their individual energies.

	\section*{A restoring soliton-energy response}
	
	\begin{figure*}[thp]
		\centering
		\includegraphics[width=\linewidth]{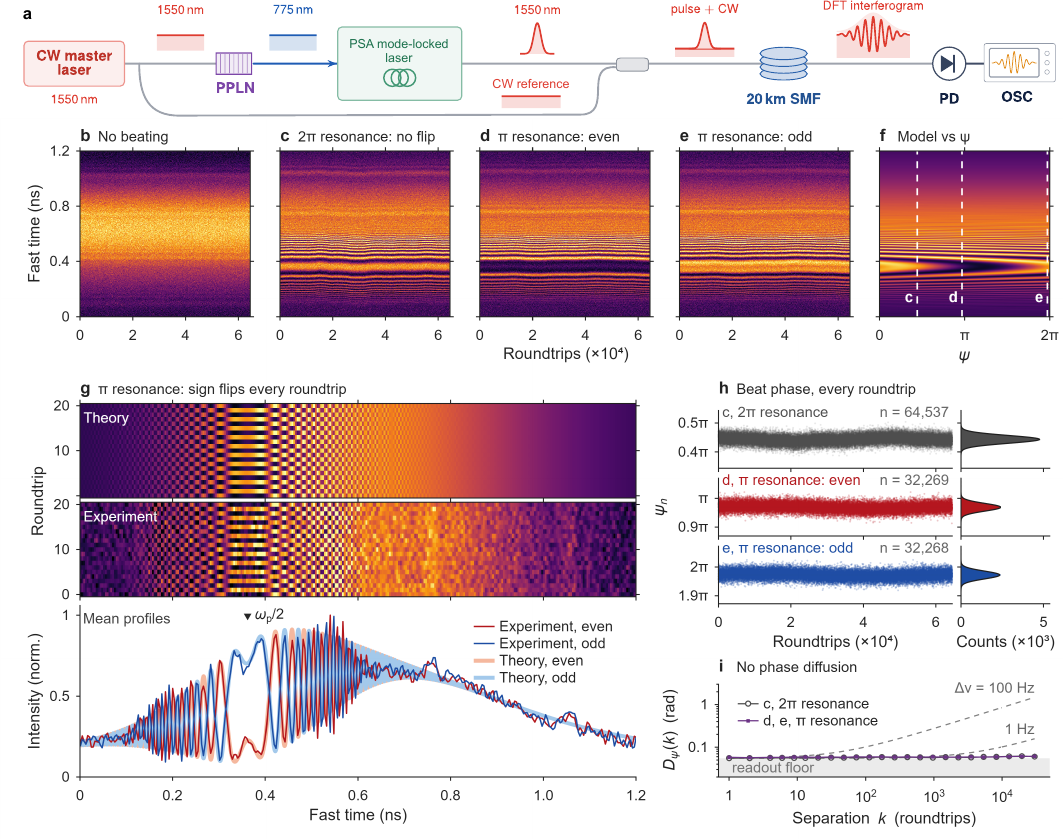}
		\caption{\textbf{Roundtrip-resolved optical phase relative to the master laser.}
			\pnl{a}, The oscillator output interferes with a tap of the 1550-nm master laser and propagates through 20~km of single-mode fiber (SMF). Dispersion stretches the pulse, producing a chirped interferogram whose phase $\psi$ tracks the soliton spectral phase at the master frequency (Methods).
			\pnl{b}--\pnl{e}, Records spanning 7.5~ms, displayed every 64 roundtrips without averaging. \pnl{b}, Reference blocked (separate acquisition). \pnl{c}, Repeating interferogram on the $2\pi$ resonance. \pnl{d},\pnl{e}, Complementary even- and odd-roundtrip interferograms on the $\pi$ resonance.
			\pnl{f}, Calculated interferogram versus relative optical phase $\psi$; dashed lines mark independently fitted mean phases.
			\pnl{g}, Twenty-one consecutive roundtrips on the $\pi$ resonance, comparing the experiment with the interferometric model. Below, the corresponding even- and odd-roundtrip mean profiles. $\blacktriangledown$, master frequency $\omega_p/2$ at the fringe waist.
			\pnl{h}, Optical phases $\psi_n$ independently fitted for individual roundtrips, with phase distributions for the repeating and alternating sequences.
			\pnl{i}, Rms phase difference $D_\psi(k)$ versus roundtrip separation $k$, after subtracting the deterministic 0 or $\pi$ advance. Shading, readout floor $D_0$ estimated by propagating measured detection noise through the fitting procedure; dashed lines, Wiener-diffusion expectations for relative linewidths of 1 and 100~Hz (Methods).
			CW, continuous wave; OSC, oscilloscope; PD, photodiode; PPLN, periodically poled lithium niobate; PSA, phase-sensitive amplification; SMF, single-mode fiber.}
		\label{fig:phase}
	\end{figure*}

	We next characterize the properties of an isolated soliton. A slow forward detuning scan prepares the single-soliton state (Supplementary Note~1), which is then followed during a backward scan over a broader detuning range (Fig.~\ref{fig:soliton}a). The measured spectral evolution, including the progressive broadening of the soliton envelope, is closely reproduced by mean-field PDNLSE simulations incorporating parametric-pump depletion, erbium gain and cavity filtering (Fig.~\ref{fig:soliton}b,c; Methods). Independently recorded radiofrequency harmonics are spaced by the 8.605-MHz cavity repetition rate (Fig.~\ref{fig:soliton}d). Model-assisted spectral analysis yields transform-limited sech$^2$-equivalent durations decreasing from approximately 2.4 to 1.5~ps and spectral redshifts of 0.29--0.58~nm (Fig.~\ref{fig:soliton}e,f).
	
	The asymmetric spectrum reveals a blue dispersive wave arising from higher-order cavity dispersion, dominated here by the spectral phase of the bandpass filter, and a red counterpart generated by parametric mixing (Fig.~\ref{fig:soliton}a,c). Additional narrow spectral features are consistent with non-flat cavity transmission, as suggested by the independently acquired no-soliton spectrum (grey curve in Fig.~\ref{fig:soliton}c). 
	The bandpass filter shapes the observed spectral structure and, in the model, suppresses higher-order parametric resonance, but is not essential for stable soliton formation, as demonstrated numerically at lower parametric drive (Supplementary Note~5).
	
	To examine pulse-energy selection, we perturb stationary model solutions at four detunings along a soliton-like amplitude--duration family preserving $P_{\rm peak}T_0^2$ (Fig.~\ref{fig:soliton}g; Methods). At the three larger detunings, the calculated one-roundtrip energy change is negative below a lower crossing, positive over an intermediate interval and negative again above a stationary energy $E_*$. This upper crossing therefore defines a locally restoring energy response, with $E_*$ increasing from approximately 40 to 49~pJ across the four detunings. Linearization shows that one modeled roundtrip opposes approximately 23--29\% of a small energy deviation along this perturbation family.
	Freezing the erbium gain at its stationary value changes the local restoring slope by less than 3\% (Methods), ruling out shared-gain saturation as its dominant origin.

	\section*{Phase coherence with the master laser}
	
	To test whether the soliton retains a defined optical phase relative to the master laser, the oscillator output is combined with a tap of the 1550-nm master laser, whose frequency-doubled field pumps the parametric amplifier, and propagated through 20~km of dispersive fiber (Fig.~\ref{fig:phase}a). The accumulated dispersion maps the optical spectrum onto a stretched temporal waveform. With the reference blocked, the measured intensity profile resembles the soliton spectral envelope independently recorded in Fig.~\ref{fig:soliton}c (Fig.~\ref{fig:phase}b).
    
	Restoring the reference reveals two distinct interferometric behaviours. The $\pi$-periodic PSA gain allows two symmetry-related field states $A$ and $-A$ (Fig.~\ref{fig:concept}e), supporting both sign-preserving and sign-alternating cavity resonances. On the $2\pi$ resonance, the chirped interference pattern repeats every roundtrip (Fig.~\ref{fig:phase}c). On the $\pi$ resonance, the field changes sign at each roundtrip, $A_{n+1}\simeq-A_n$, producing complementary even- and odd-roundtrip patterns (Fig.~\ref{fig:phase}d,e). The fringe phase encodes the soliton optical phase relative to the master, up to an uncalibrated reference-path offset. An interferometric model reproduces the observed patterns (Fig.~\ref{fig:phase}f,g), allowing the phase of each unaveraged roundtrip to be extracted independently, without imposing the expected $\pi$ difference (Methods).

    The extracted phases confirm this distinction over 64,537 consecutive roundtrips (7.5 ms). The $2\pi$ resonance maintains a single phase class, whereas the $\pi$ resonance alternates between two classes separated by $\pi$ (Fig.~\ref{fig:phase}h), with no resolved phase slips. To quantify phase stability, we calculate $D_\psi(k)$, the root-mean-square phase difference between roundtrips separated by $k$, after removing the expected 0 or $\pi$ phase advance. Even for a perfectly stable optical phase, detection noise produces a finite difference. We estimate this measurement floor, $D_0\simeq0.055$ rad, by propagating the measured detector noise through the phase-extraction procedure (Methods). The observed $D_\psi$ remains close to this floor across the measurement window (Fig.~\ref{fig:phase}i), rather than increasing as predicted for Wiener phase diffusion with relative linewidths of 1 and 100 Hz. At a separation of 3.5 ms, $D_\psi\simeq0.061$ rad (Methods).
    These measurements establish master-referenced optical-phase coherence with the expected binary symmetry, confirming phase-sensitive mode locking.

	\section*{Discussion}

    Phase-sensitive mode locking makes the external optical reference part of the pulse-selection mechanism itself. Kerr propagation encodes pulse power into optical phase, which phase-sensitive amplification converts into gain or loss relative to the coherent pump. In our oscillator, this feedback sustains parametrically driven Kerr solitons, with a locally restoring energy response. Saturation of the auxiliary erbium gain plays a distinct role, governing competition among solitons and limiting the number that survive. Building on earlier theories of parametric pulse formation~\cite{Longhi1995} and Kerr--PSA mode locking~\cite{Kutz2008}, our results establish this coherent pulse-selection mechanism in a continuously pumped oscillator, where no synchronous pulse-arrival slots are imposed~\cite{Kobayashi2015,Obrzud2017}.
    
    The same parametric interaction also provides an optical phase reference without an additive coherent drive at the signal frequency. Roundtrip-resolved interferometry reveals master-referenced coherence with a repeating optical phase on the $2\pi$ resonance and alternating phases separated by $\pi$ on the $\pi$ resonance. These two resonance classes have distinct frequency-comb structures: in a stationary comb, the $2\pi$ class places its teeth at the master frequency plus integer multiples of $f_{\rm rep}$, whereas the $\pi$ class shifts them by $f_{\rm rep}/2$ (Supplementary Note 7). Stabilizing the master laser to an optical standard would therefore anchor one optical degree of freedom of the pulse train through the pulse-forming interaction itself.
    
    The two symmetry-related field states, $A$ and $-A$, also provide a natural binary variable for parametric-oscillator networks~\cite{Marandi2014}, potentially extendable to integrated parametric platforms~\cite{Bruch2021,Lu2026}. Cavity detuning offers a complementary degree of control: small changes produce substantial shifts of the soliton spectral centre. Together with pump-referenced optical coherence, this sensitivity points towards spectrally agile ultrafast sources whose spectral envelope can be tuned relative to an external optical reference.

	
	\renewcommand{\refname}{References}
	\bibliographystyle{naturemag-numbered}
	\bibliography{references2}

@article{haus2000mode,
  author  = {Haus, Hermann A.},
  title   = {Mode-locking of lasers},
  journal = {IEEE J. Sel. Top. Quantum Electron.},
  volume  = {6},
  pages   = {1173--1185},
  year    = {2000},
  doi     = {10.1109/2944.902165},
  refnumber = {1}
}

@article{Keller2003,
  author  = {Keller, Ursula},
  title   = {Recent developments in compact ultrafast lasers},
  journal = {Nature},
  volume  = {424},
  pages   = {831--838},
  year    = {2003},
  doi     = {10.1038/nature01938},
  refnumber = {2}
}

@article{udem2002optical,
  author  = {Udem, Th. and Holzwarth, R. and H{\"a}nsch, T. W.},
  title   = {Optical frequency metrology},
  journal = {Nature},
  volume  = {416},
  pages   = {233--237},
  year    = {2002},
  doi     = {10.1038/416233a},
  refnumber = {5}
}

@article{Diddams2020,
  author  = {Diddams, Scott A. and Vahala, Kerry and Udem, Thomas},
  title   = {Optical frequency combs: coherently uniting the electromagnetic spectrum},
  journal = {Science},
  volume  = {369},
  pages   = {eaay3676},
  year    = {2020},
  doi     = {10.1126/science.aay3676},
  refnumber = {6}
}

@article{Telle1999,
  author  = {Telle, H. R. and Steinmeyer, G. and Dunlop, A. E. and Stenger, J. and Sutter, D. H. and Keller, U.},
  title   = {Carrier-envelope offset phase control: a novel concept for absolute optical frequency measurement and ultrashort pulse generation},
  journal = {Appl. Phys. B},
  volume  = {69},
  pages   = {327--332},
  year    = {1999},
  doi     = {10.1007/s003400050813},
  refnumber = {9}
}

@article{Jones2000,
  author  = {Jones, David J. and Diddams, Scott A. and Ranka, Jinendra K. and Stentz, Andrew and Windeler, Robert S. and Hall, John L. and Cundiff, Steven T.},
  title   = {Carrier-envelope phase control of femtosecond mode-locked lasers and direct optical frequency synthesis},
  journal = {Science},
  volume  = {288},
  pages   = {635--639},
  year    = {2000},
  doi     = {10.1126/science.288.5466.635},
  refnumber = {10}
}

@article{Fortier2019,
  author  = {Fortier, Tara and Baumann, Esther},
  title   = {20 years of developments in optical frequency comb technology and applications},
  journal = {Commun. Phys.},
  volume  = {2},
  pages   = {153},
  year    = {2019},
  doi     = {10.1038/s42005-019-0249-y},
  refnumber = {11}
}

@article{Longhi1995,
  author  = {Longhi, Stefano},
  title   = {Ultrashort-pulse generation in degenerate optical parametric oscillators},
  journal = {Opt. Lett.},
  volume  = {20},
  pages   = {695--697},
  year    = {1995},
  doi     = {10.1364/OL.20.000695},
  refnumber = {12}
}

@article{Kutz2008,
  author  = {Kutz, J. Nathan},
  title   = {Passive mode-locking using phase-sensitive amplification},
  journal = {Phys. Rev. A},
  volume  = {78},
  pages   = {013845},
  year    = {2008},
  doi     = {10.1103/PhysRevA.78.013845},
  refnumber = {13}
}

@article{Fermann2013,
  author  = {Fermann, Martin E. and Hartl, Ingmar},
  title   = {Ultrafast fibre lasers},
  journal = {Nat. Photon.},
  volume  = {7},
  pages   = {868--874},
  year    = {2013},
  doi     = {10.1038/nphoton.2013.280},
  refnumber = {3}
}

@article{sun2019mode,
  author  = {Sun, Yifan and Combri{\'e}, Sylvain and Bretenaker, Fabien and {De Rossi}, Alfredo},
  title   = {Mode Locking of the {Hermite--Gaussian} Modes of a Nanolaser},
  journal = {Phys. Rev. Lett.},
  volume  = {123},
  number  = {23},
  pages   = {233901},
  year    = {2019},
  doi     = {10.1103/PhysRevLett.123.233901},
  refnumber = {4}
}

@article{Grelu2012,
  author  = {Grelu, Philippe and Akhmediev, Nail},
  title   = {Dissipative solitons for mode-locked lasers},
  journal = {Nat. Photon.},
  volume  = {6},
  pages   = {84--92},
  year    = {2012},
  doi     = {10.1038/nphoton.2011.345},
  refnumber = {14}
}

@article{Keller1996,
  author  = {Keller, U. and Weingarten, K. J. and K{\"a}rtner, F. X. and Kopf, D. and Braun, B. and Jung, I. D. and Fluck, R. and H{\"o}nninger, C. and Matuschek, N. and {Aus der Au}, J.},
  title   = {Semiconductor saturable absorber mirrors ({SESAM}'s) for femtosecond to nanosecond pulse generation in solid-state lasers},
  journal = {IEEE J. Sel. Top. Quantum Electron.},
  volume  = {2},
  pages   = {435--453},
  year    = {1996},
  doi     = {10.1109/2944.571743},
  refnumber = {15}
}

@article{Spence1991,
  author  = {Spence, D. E. and Kean, P. N. and Sibbett, W.},
  title   = {60-fsec pulse generation from a self-mode-locked {Ti:sapphire} laser},
  journal = {Opt. Lett.},
  volume  = {16},
  pages   = {42--44},
  year    = {1991},
  doi     = {10.1364/OL.16.000042},
  refnumber = {16}
}

@article{Hofer1991,
  author  = {Hofer, M. and Fermann, M. E. and Haberl, F. and Ober, M. H. and Schmidt, A. J.},
  title   = {Mode locking with cross-phase and self-phase modulation},
  journal = {Opt. Lett.},
  volume  = {16},
  pages   = {502--504},
  year    = {1991},
  doi     = {10.1364/OL.16.000502},
  refnumber = {17}
}

@article{Doran1988,
  author  = {Doran, N. J. and Wood, David},
  title   = {Nonlinear-optical loop mirror},
  journal = {Opt. Lett.},
  volume  = {13},
  pages   = {56--58},
  year    = {1988},
  doi     = {10.1364/OL.13.000056},
  refnumber = {18}
}

@article{Fermann1990,
  author  = {Fermann, M. E. and Haberl, F. and Hofer, M. and Hochreiter, H.},
  title   = {Nonlinear amplifying loop mirror},
  journal = {Opt. Lett.},
  volume  = {15},
  pages   = {752--754},
  year    = {1990},
  doi     = {10.1364/OL.15.000752},
  refnumber = {19}
}

@article{Baltuska2003,
  author  = {Baltu{\v{s}}ka, A. and Udem, Th. and Uiberacker, M. and Hentschel, M. and Goulielmakis, E. and Gohle, Ch. and Holzwarth, R. and Yakovlev, V. S. and Scrinzi, A. and H{\"a}nsch, T. W. and Krausz, F.},
  title   = {Attosecond control of electronic processes by intense light fields},
  journal = {Nature},
  volume  = {421},
  pages   = {611--615},
  year    = {2003},
  doi     = {10.1038/nature01414},
  refnumber = {7}
}

@article{Fortier2011,
  author  = {Fortier, T. M. and Kirchner, M. S. and Quinlan, F. and Taylor, J. and Bergquist, J. C. and Rosenband, T. and Lemke, N. and Ludlow, A. and Jiang, Y. and Oates, C. W. and Diddams, S. A.},
  title   = {Generation of ultrastable microwaves via optical frequency division},
  journal = {Nat. Photon.},
  volume  = {5},
  pages   = {425--429},
  year    = {2011},
  doi     = {10.1038/nphoton.2011.121},
  refnumber = {8}
}

@article{Leo2010,
  author  = {Leo, Fran{\c{c}}ois and Coen, St{\'e}phane and Kockaert, Pascal and Gorza, Simon-Pierre and Emplit, Philippe and Haelterman, Marc},
  title   = {Temporal cavity solitons in one-dimensional {Kerr} media as bits in an all-optical buffer},
  journal = {Nat. Photon.},
  volume  = {4},
  pages   = {471--476},
  year    = {2010},
  doi     = {10.1038/nphoton.2010.120},
  refnumber = {20}
}

@article{Herr2014,
  author  = {Herr, T. and Brasch, V. and Jost, J. D. and Wang, C. Y. and Kondratiev, N. M. and Gorodetsky, M. L. and Kippenberg, T. J.},
  title   = {Temporal solitons in optical microresonators},
  journal = {Nat. Photon.},
  volume  = {8},
  pages   = {145--152},
  year    = {2014},
  doi     = {10.1038/nphoton.2013.343},
  refnumber = {21}
}

@article{EnglebertActive2021,
  author  = {Englebert, Nicolas and {Mas Arab{\'i}}, Carlos and Parra-Rivas, Pedro and Gorza, Simon-Pierre and Leo, Fran{\c{c}}ois},
  title   = {Temporal solitons in a coherently driven active resonator},
  journal = {Nat. Photon.},
  volume  = {15},
  pages   = {536--541},
  year    = {2021},
  doi     = {10.1038/s41566-021-00807-w},
  refnumber = {22}
}

@article{EnglebertContrast2023,
  author  = {Englebert, Nicolas and {Mas Arab{\'i}}, Carlos and Gorza, Simon-Pierre and Leo, Fran{\c{c}}ois},
  title   = {High peak-to-background-ratio solitons in a coherently driven active fiber cavity},
  journal = {APL Photonics},
  volume  = {8},
  number  = {12},
  pages   = {120802},
  year    = {2023},
  doi     = {10.1063/5.0159693},
  refnumber = {27}
}

@article{Caves1982,
  author  = {Caves, Carlton M.},
  title   = {Quantum limits on noise in linear amplifiers},
  journal = {Phys. Rev. D},
  volume  = {26},
  pages   = {1817--1839},
  year    = {1982},
  doi     = {10.1103/PhysRevD.26.1817},
  refnumber = {28}
}

@article{Marandi2014,
  author  = {Marandi, Alireza and Wang, Zhe and Takata, Kenta and Byer, Robert L. and Yamamoto, Yoshihisa},
  title   = {Network of time-multiplexed optical parametric oscillators as a coherent {Ising} machine},
  journal = {Nat. Photon.},
  volume  = {8},
  pages   = {937--942},
  year    = {2014},
  doi     = {10.1038/nphoton.2014.249},
  refnumber = {43}
}

@article{Wong2010,
  author  = {Wong, Samuel T. and Vodopyanov, Konstantin L. and Byer, Robert L.},
  title   = {Self-phase-locked divide-by-2 optical parametric oscillator as a broadband frequency comb source},
  journal = {J. Opt. Soc. Am. B},
  volume  = {27},
  pages   = {876--882},
  year    = {2010},
  doi     = {10.1364/JOSAB.27.000876},
  refnumber = {29}
}

@article{Marandi2012Coherence,
  author  = {Marandi, Alireza and Leindecker, Nick C. and Pervak, Vladimir and Byer, Robert L. and Vodopyanov, Konstantin L.},
  title   = {Coherence properties of a broadband femtosecond mid-{IR} optical parametric oscillator operating at degeneracy},
  journal = {Opt. Express},
  volume  = {20},
  pages   = {7255--7262},
  year    = {2012},
  doi     = {10.1364/OE.20.007255},
  refnumber = {30}
}

@article{Bruch2021,
  author  = {Bruch, Alexander W. and Liu, Xianwen and Gong, Zheng and Surya, Joshua B. and Li, Ming and Zou, Chang-Ling and Tang, Hong X.},
  title   = {{Pockels} soliton microcomb},
  journal = {Nat. Photon.},
  volume  = {15},
  pages   = {21--27},
  year    = {2021},
  doi     = {10.1038/s41566-020-00704-8},
  refnumber = {44}
}

@article{Moille2024,
  author  = {Moille, Gr{\'e}gory and Leonhardt, Miriam and Paligora, David and Englebert, Nicolas and Leo, Fran{\c{c}}ois and Fatome, Julien and Srinivasan, Kartik and Erkintalo, Miro},
  title   = {Parametrically driven pure-{Kerr} temporal solitons in a chip-integrated microcavity},
  journal = {Nat. Photon.},
  volume  = {18},
  pages   = {617--624},
  year    = {2024},
  doi     = {10.1038/s41566-024-01401-6},
  refnumber = {32}
}

@article{sunParametricallyDrivenKerr2026a,
  author  = {Sun, Yifan and Dupont, Cl{\'e}ment and Kossi Akakpo, Edem and De Lucia, Francesco and Semaan, Georges and Gorza, Simon-Pierre and Leo, Fran{\c{c}}ois},
  title   = {Parametrically driven {Kerr} temporal soliton crystals},
  journal = {J. Phys. Photonics},
  volume  = {8},
  number  = {3},
  pages   = {03LT01},
  year    = {2026},
  doi     = {10.1088/2515-7647/ae79bf},
  refnumber = {33}
}

@article{Englebert2021,
  author  = {Englebert, Nicolas and De Lucia, Francesco and Parra-Rivas, Pedro and {Mas Arab{\'i}}, Carlos and Sazio, Pier-John and Gorza, Simon-Pierre and Leo, Fran{\c{c}}ois},
  title   = {Parametrically driven {Kerr} cavity solitons},
  journal = {Nat. Photon.},
  volume  = {15},
  pages   = {857--861},
  year    = {2021},
  doi     = {10.1038/s41566-021-00858-z},
  refnumber = {31}
}

@article{Nie2020OL,
  author  = {Nie, Mingming and Huang, Shu-Wei},
  title   = {Quadratic soliton mode-locked degenerate optical parametric oscillator},
  journal = {Opt. Lett.},
  volume  = {45},
  pages   = {2311--2314},
  year    = {2020},
  doi     = {10.1364/OL.389568},
  refnumber = {36}
}

@article{Nie2020PRApplied,
  author  = {Nie, Mingming and Huang, Shu-Wei},
  title   = {Quadratic solitons in singly resonant degenerate optical parametric oscillators},
  journal = {Phys. Rev. Appl.},
  volume  = {13},
  pages   = {044046},
  year    = {2020},
  doi     = {10.1103/PhysRevApplied.13.044046},
  refnumber = {37}
}

@misc{Musgrave2026,
  author       = {Musgrave, Jonathan and Nie, Mingming and Huang, Shu-Wei},
  title        = {Dissipative quadratic soliton mode-locked optical parametric oscillator},
  howpublished = {Preprint at \url{https://arxiv.org/abs/2604.03474}},
  year         = {2026},
  refnumber = {38}
}

@article{Tang2005,
  author  = {Tang, D. Y. and Zhao, L. M. and Zhao, B. and Liu, A. Q.},
  title   = {Mechanism of multisoliton formation and soliton energy quantization in passively mode-locked fiber lasers},
  journal = {Phys. Rev. A},
  volume  = {72},
  pages   = {043816},
  year    = {2005},
  doi     = {10.1103/PhysRevA.72.043816},
  refnumber = {40}
}

@article{Moille2023,
  author  = {Moille, Gr{\'e}gory and Stone, Jordan and Chojnacky, Michal and Shrestha, Rahul and Javid, Usman A. and Menyuk, Curtis and Srinivasan, Kartik},
  title   = {{Kerr}-induced synchronization of a cavity soliton to an optical reference},
  journal = {Nature},
  volume  = {624},
  pages   = {267--274},
  year    = {2023},
  doi     = {10.1038/s41586-023-06730-0},
  refnumber = {26}
}

@article{Kobayashi2015,
  author  = {Kobayashi, Yohei and Torizuka, Kenji and Marandi, Alireza and Byer, Robert L. and McCracken, Richard A. and Zhang, Zhaowei and Reid, Derryck T.},
  title   = {Femtosecond optical parametric oscillator frequency combs},
  journal = {J. Opt.},
  volume  = {17},
  pages   = {094010},
  year    = {2015},
  doi     = {10.1088/2040-8978/17/9/094010},
  refnumber = {41}
}

@article{Obrzud2017,
  author  = {Obrzud, Ewelina and Lecomte, Steve and Herr, Tobias},
  title   = {Temporal solitons in microresonators driven by optical pulses},
  journal = {Nat. Photon.},
  volume  = {11},
  pages   = {600--607},
  year    = {2017},
  doi     = {10.1038/nphoton.2017.140},
  refnumber = {42}
}

@article{Lu2026,
  author  = {Lu, Xiyuan and Gray, Robert M. and Stone, Jordan and Zhou, Selina and Englebert, Nicolas and Marandi, Alireza and Srinivasan, Kartik},
  title   = {Photonic integrated circuit optical parametric oscillators},
  journal = {Optica},
  volume  = {13},
  pages   = {11--30},
  year    = {2026},
  doi     = {10.1364/OPTICA.572694},
  refnumber = {45}
}

@article{Bosenberg1996,
  author  = {Bosenberg, Walter R. and Drobshoff, Alexander and Alexander, Jason I. and Myers, Lawrence E. and Byer, Robert L.},
  title   = {Continuous-wave singly resonant optical parametric oscillator based on periodically poled {LiNbO}$_3$},
  journal = {Opt. Lett.},
  volume  = {21},
  pages   = {713--715},
  year    = {1996},
  doi     = {10.1364/OL.21.000713},
  refnumber = {46}
}

@article{Drever1983,
  author  = {Drever, R. W. P. and Hall, J. L. and Kowalski, F. V. and Hough, J. and Ford, G. M. and Munley, A. J. and Ward, H.},
  title   = {Laser phase and frequency stabilization using an optical resonator},
  journal = {Appl. Phys. B},
  volume  = {31},
  pages   = {97--105},
  year    = {1983},
  doi     = {10.1007/BF00702605},
  refnumber = {47}
}

@article{Haelterman1992,
  author  = {Haelterman, M. and Trillo, S. and Wabnitz, S.},
  title   = {Dissipative modulation instability in a nonlinear dispersive ring cavity},
  journal = {Opt. Commun.},
  volume  = {91},
  pages   = {401--407},
  year    = {1992},
  doi     = {10.1016/0030-4018(92)90367-Z},
  refnumber = {50}
}

@article{CoenModeling2013,
  author  = {Coen, St{\'e}phane and Randle, Hamish G. and Sylvestre, Thibaut and Erkintalo, Miro},
  title   = {Modeling of octave-spanning {Kerr} frequency combs using a generalized mean-field {Lugiato--Lefever} model},
  journal = {Opt. Lett.},
  volume  = {38},
  pages   = {37--39},
  year    = {2013},
  doi     = {10.1364/OL.38.000037},
  refnumber = {51}
}

@article{Miles1984,
  author  = {Miles, John W.},
  title   = {Parametrically excited solitary waves},
  journal = {J. Fluid Mech.},
  volume  = {148},
  pages   = {451--460},
  year    = {1984},
  doi     = {10.1017/S0022112084002433},
  refnumber = {48}
}

@article{Barashenkov1991,
  author  = {Barashenkov, I. V. and Bogdan, M. M. and Korobov, V. I.},
  title   = {Stability diagram of the phase-locked solitons in the parametrically driven, damped nonlinear {Schr{\"o}dinger} equation},
  journal = {Europhys. Lett.},
  volume  = {15},
  pages   = {113--118},
  year    = {1991},
  doi     = {10.1209/0295-5075/15/2/001},
  refnumber = {49}
}

@article{MasArabi2023,
  author  = {{Mas Arab{\'i}}, Carlos and Englebert, Nicolas and Parra-Rivas, Pedro and Gorza, Simon-Pierre and Leo, Fran{\c{c}}ois},
  title   = {Depletion-limited {Kerr} solitons in singly resonant optical parametric oscillators},
  journal = {Opt. Lett.},
  volume  = {48},
  number  = {7},
  pages   = {1950--1953},
  year    = {2023},
  doi     = {10.1364/OL.483871},
  refnumber = {39}
}

@article{Blow1989,
  author  = {Blow, K. J. and Wood, D.},
  title   = {Theoretical description of transient stimulated {Raman} scattering in optical fibers},
  journal = {IEEE J. Quantum Electron.},
  volume  = {25},
  pages   = {2665--2673},
  year    = {1989},
  doi     = {10.1109/3.40655},
  refnumber = {52}
}

@article{Mahjoubfar2017,
  author  = {Mahjoubfar, Ata and Churkin, Dmitry V. and Barland, St{\'e}phane and Broderick, Neil and Turitsyn, Sergei K. and Jalali, Bahram},
  title   = {Time stretch and its applications},
  journal = {Nat. Photon.},
  volume  = {11},
  pages   = {341--351},
  year    = {2017},
  doi     = {10.1038/nphoton.2017.76},
  refnumber = {53}
}

@article{Englebert2026,
  author  = {Englebert, Nicolas and Gray, Robert M. and Ledezma, Luis and Sekine, Ryoto and Zacharias, Thomas and Ramesh, Rithvik and Gutierrez, Benjamin K. and Parra-Rivas, Pedro and Marandi, Alireza},
  title   = {Topological soliton frequency comb in nanophotonic lithium niobate},
  journal = {Nature},
  volume  = {652},
  pages   = {76--81},
  year    = {2026},
  doi     = {10.1038/s41586-026-10292-2},
  refnumber = {34}
}

@article{semaanCavitySolitonsUltrahigheffectiveQ2026,
  title = {Cavity Solitons in an Ultrahigh-Effective-{{Q}} Fiber Resonator Enabled by Raman Gain},
  author = {Semaan, Georges and Sun, Yifan and Englebert, Nicolas and Gorza, Simon-Pierre and Leo, Fran{\c c}ois},
  year = 2026,
  month = aug,
  journal = {APL Photonics},
  volume = {11},
  number = {8},
  pages = {81301},
  issn = {2378-0967},
  doi = {10.1063/5.0340179},
  urldate = {2026-08-11},
  language = {en},
  refnumber = {23}
}

@article{Liang2026,
  author = {Liang, Yupei and Zhang, Dingfu and Guo, Genhong
            and Zhang, Hao and Gao, Muyan and Rao, Yunjiang
            and Wong, Chee Wei and Tan, Teng and Yao, Baicheng},
  title = {Dissipative {Kerr} solitons in an active microcavity},
  journal = {Laser Photonics Rev.},
  volume = {20},
  number = {14},
  pages = {e02552},
  year = {2026},
  doi = {10.1002/lpor.202502552},
  refnumber = {24}
}

@article{Kazakov2025,
  author = {Kazakov, Dmitry and Letsou, Theodore P.
            and Piccardo, Marco and Columbo, Lorenzo L.
            and Brambilla, Massimo and Prati, Franco
            and Dal Cin, Sandro and Beiser, Maximilian
            and Opa{\v{c}}ak, Nikola and Ratra, Pawan
            and Pushkarsky, Michael and Caffey, David
            and Day, Timothy and Lugiato, Luigi A.
            and Schwarz, Benedikt and Capasso, Federico},
  title = {Driven bright solitons on a mid-infrared laser chip},
  journal = {Nature},
  volume = {641},
  pages = {83--89},
  year = {2025},
  doi = {10.1038/s41586-025-08853-y},
  refnumber = {25}
}

@article{Moille2026,
  author  = {Moille, Gr{\'e}gory and Shandilya, Pradyoth and Stone, Jordan and Beard, River and Ou, Shao-Chien and Li, Zongda and Harrington, Mark and Liu, Kaikai and Rockmore, Robert and Menyuk, Curtis R. and Blumenthal, Daniel J. and Krzyzewski, Sean P. and Erkintalo, Miro and Srinivasan, Kartik},
  title   = {Self-aligned optical microcomb emerging between octave-separated lasers},
  journal = {Nature},
  year    = {2026},
  doi     = {10.1038/s41586-026-11086-2},
  refnumber = {35}
}
	
	\clearpage

	\section*{Methods}
	
	\paragraph{Oscillator and detuning control.}
	The singly resonant ring contains 23.86~m of fiber (SMF-28, component pigtails and 20~cm of erbium-doped fiber), a 5-cm fiber-coupled periodically poled lithium niobate (PPLN) waveguide (HC Photonics Corp.), a nominal 5-nm bandpass filter, polarization controllers and two 99/1 couplers. The PPLN module has a quoted transmission of 82\% at both the pump and signal wavelengths. A continuous-wave master laser near 1550~nm is amplified and frequency doubled in a separate PPLN crystal. The resulting 775-nm pump traverses the intracavity waveguide once, while only the subharmonic signal resonates, as in a singly resonant optical parametric oscillator~\cite{Bosenberg1996}. The 20-cm erbium section is pumped with 0.9~W at 980~nm and partially compensates the passive loss; with the parametric pump blocked, its gain remains below the cavity lasing threshold. Thus the gain reduces the net weak-signal attenuation without changing the cold-cavity loss. 
    A frequency-shifted, phase-modulated auxiliary field derived from the master laser counterpropagates through the resonator with polarization orthogonal to the pump and signal fields. Its detected response provides a Pound--Drever--Hall error signal~\cite{Drever1983}, which is fed back to the master laser for detuning control. Further experimental details are provided in Supplementary Note~1.
    The measured repetition frequency is $f_{\rm rep}=8.605$~MHz ($T_R=116.21$~ns). Figures~\ref{fig:selection}--\ref{fig:phase} use datasets acquired in separate sessions.

	\paragraph{Phase-sensitive gain and pulse-phase overlap.}
	In the undepleted-pump limit, the PPLN interaction transforms the signal envelope as $A_{\rm out}=\cosh(q)A_{\rm in}+i\exp(i\phi_p)\sinh(q)A_{\rm in}^{*}$, where $q=\kappa L_c\sqrt{P_p}$, $L_c=5$~cm, and $P_p$ and $\phi_p$ are the pump power and phase at the crystal input. For pulse energy $E=\int|A|^2\,d\tau$, define $\rho=|\int A^2\,d\tau|/E$ and $\Phi=\tfrac12\arg(\int A^2\,d\tau)$, with $\Phi$ defined modulo $\pi$. The resulting fractional energy gain is
	\(
	G=\frac{E_{\rm out}}{E_{\rm in}}=\cosh(2q)+\rho\sinh(2q)\cos(2\varphi),
	\)
	where $\varphi=\Phi-\frac{\phi_p}{2}-\frac{\pi}{4}$.
	Maximum gain occurs at $\varphi=0$ modulo $\pi$, so $A$ and $-A$ have the same gain. The overlap $\rho$ quantifies the reduction in phase-sensitive gain contrast caused by temporal phase variations across the pulse: it equals one for a uniform temporal phase and decreases with chirp or carrier-frequency displacement (Supplementary Table~S3). 
    For the fitted cavity solitons, the pulse-window overlap is approximately 0.77--0.81, rising to 0.92--0.98 after removal of the carrier-frequency shift (Supplementary Note~2). The ideal fundamental-soliton phase advance, $\gamma PL_f/2$ over fiber length $L_f$ at peak power $P$, motivates the Kerr phase-encoding picture but is not imposed on the dissipative solutions. Figure~\ref{fig:concept}f normalized single-pass illustrations of the Kerr–PSA discriminator. The gain convention and pump-depletion controls are detailed in Supplementary Note~2.
		
	\paragraph{Mean-field cavity model.}
	Figures~\ref{fig:selection} and~\ref{fig:soliton} are calculated using a mean-field parametrically driven nonlinear Schr\"odinger equation (PDNLSE)~\cite{Miles1984,Barashenkov1991} in a cavity formulation~\cite{Haelterman1992,CoenModeling2013}. The model includes logarithmic loss and detuning, second- and third-order dispersion, Kerr nonlinearity, a first-moment Raman term~\cite{Blow1989}, phase-sensitive parametric driving with pump depletion, saturable erbium gain and the filter's spectral response. Pump depletion is represented by a first-order response kernel evaluated from the instantaneous signal, with pump--signal walk-off of 300~ps~m$^{-1}$~\cite{MasArabi2023}; the 5-cm crystal therefore gives approximately 15~ps of walk-off. The depletion kernel is updated at every Runge--Kutta stage. 
    Equations and normalization are given in Supplementary Note~3.
	
	\paragraph{Fitted parameters and spectral filtering.}
	The mean-field parameter set used for Fig.~\ref{fig:soliton} has $\kappa=2.7~\mathrm{W}^{-1/2}\mathrm{m}^{-1}$, $\beta_2=-21.17~\mathrm{ps}^2\mathrm{km}^{-1}$, $\gamma=1.22~\mathrm{W}^{-1}\mathrm{km}^{-1}$, $\beta_3=0.12~\mathrm{ps}^3\mathrm{km}^{-1}$ and a Raman time of 3~fs. The model pump power is 0.773~W. The logarithmic cold-cavity loss is $\Lambda_{\rm cold}=0.55492$; the unsaturated erbium gain is $g_0=0.49075$, and the saturated gain is $g=g_0/(1+P_{\rm av}/P_{\rm sat})$, with $P_{\rm sat}=27$~mW. Here $P_{\rm av}$ is the circulating energy in the numerical window, plus any background contribution from the rest of the physical roundtrip, divided by $T_R$ (116~ns in the simulations). The quantity $P_{\rm sat}$ is defined by halving the model's logarithmic erbium gain, not by parametric-pump saturation or an output 3-dB power specification. The net unsaturated attenuation without parametric pumping is $\Lambda_{\rm uns}=\Lambda_{\rm cold}-g_0=0.06417$, giving $\mu\equiv2q/\Lambda_{\rm uns}=3.70$. The corresponding formal loss-budget ratio $ \mathcal{F}_\mathrm{eff}=2\pi/\Lambda_{\rm uns}\simeq97.9$ is described as an effective-finesse estimate.
	
	The model was adjusted to reproduce the measured spectral evolution. In particular, $\beta_2$, $\gamma$, $\Lambda_{\rm cold}$ and the pump power are effective fitted values; $\kappa$ is held at $2.7~\mathrm{W}^{-1/2}\mathrm{m}^{-1}$ and $P_{\rm sat}$ at 27~mW, a value also used in Fig.~\ref{fig:selection}, after which $\mu$ was selected by a one-dimensional scan; the erbium parameters are phenomenological. The filter amplitude was measured, but its phase was reconstructed using a minimum-phase assumption. In the model, this phase shapes higher-order dispersive features and the accessible soliton branch. 
    Filter-free simulations also support stable localized states, primarily at lower parametric drive and under different spectral conditions, whereas spectral filtering extends the stability region to include the fitted operating regime at $\mu\simeq3.70$ (Supplementary Note~5). Parameter choices and identifiability are discussed in Supplementary Note~3 and Supplementary Table~S5.

	\paragraph{Experimental population record.}
	For Fig.~\ref{fig:selection}a,b, the cavity was swept rapidly forward with the Pound--Drever--Hall loop disengaged. A 70-GHz photodiode (Finisar XPDV3320R) and an 80-GHz oscilloscope (Keysight UXR80) acquired 25,815 consecutive roundtrips (approximately 3.0~ms). Each 116.21-ns interval forms one row of the space--time map.
	
	\paragraph{Erbium saturation and population simulations.}
	Figure~\ref{fig:selection} uses a cavity parameter on a 5-ns grid of 65,536 samples. Within each comparison, only $P_{\rm sat}$ is varied. The detuning is scanned from $-0.30$ to $+1.00$~rad at $6.5\times10^{-5}$~rad per roundtrip, with pulse counts every 20 roundtrips. Seven noise-seeded runs were computed for each of $P_{\rm sat}=18$, 27 and 100~mW and for unsaturated erbium gain. For $N$ comparable, widely separated pulses of energy $E_s$, $P_{\rm av}\simeq NE_s/T_R$, so their common gain equals the single-pulse saturation law after $P_{\rm sat}\rightarrow P_{\rm sat}/N$. The independent one-soliton loss boundary $\delta_1(P)$ was obtained by continuing an analytic soliton initialized at $\delta=0.08$~rad in a 250-ps window at the same scan rate, without fitting this boundary to the population data. Under the pulse-tracking criterion, 221 of 251 losses at the onset of a collapse occur within 0.003~rad of the boundary. Closely bound pairs are the main exceptions because local pump depletion violates the independent-pulse approximation (Supplementary Note~4).
	
	\paragraph{Spectral scan and detection.}
	For Fig.~\ref{fig:soliton}, a single soliton was prepared with a slow forward scan and followed during a backward scan. The acquisition contains 160 spectra over a modulation-frequency interval of 173.60 to 172.72~MHz, with approximately 5.53-kHz frequency steps, 2~s per trace and 0.5~s settling time per step. The recorded parametric-pump power at the output was 0.5~W. The plotted detuning is calculated as $\delta=2\pi(f-f_{\rm cw})/\mathrm{FSR}-D$, with $f_{\rm cw}=172.95$~MHz, $\mathrm{FSR}=8.605$~MHz and $D=0.0460$~rad (Supplementary Note~1). The reference offset was selected to align the measured and simulated soliton durations and redshifts  (Supplementary Note~5). 

	Spectra were acquired with a Yokogawa AQ6370 analyser from 1,545 to 1,555~nm at 0.01-nm sampling and nominal 0.02-nm resolution. Simulated spectra were convolved with a Gaussian 0.02-nm-FWHM intensity response, resampled to the experimental grid and displayed with a measured noise-floor estimate. The radiofrequency spectrum in Fig.~\ref{fig:soliton}d was acquired separately using a 25-GHz photodiode and a Rohde \& Schwarz FSP analyser. The 8.605-MHz harmonic spacing establishes the pulse repetition period.
	
	\paragraph{Spectral-envelope fits and background assessment.}
	The experimental and simulated spectra in Fig.~\ref{fig:soliton}c are normalized separately to their envelope maxima. 
    For Fig.~\ref{fig:soliton}e,f, the measured spectral envelopes are fitted with model-derived templates to estimate the pulse duration and spectral redshift. The fitted bandwidth $\Delta\nu$ gives a transform-limited sech$^2$-equivalent duration, $\Delta t_{\rm eq}=0.3148/\Delta\nu$, while the redshift is measured relative to 1,549.284~nm. Shading indicates half the spread across three fitting weightings. 
    The grey no-soliton trace in Fig.~\ref{fig:soliton}c comes from a separate acquisition and serves to identify stationary spectral structures, not to subtract a background (Supplementary Note~5).
	
	\paragraph{Numerical continuation and energy perturbations.}
	The Fig.~\ref{fig:soliton} simulations use a 2-ns computational window with 16,384 samples, symmetric split steps of 0.1 roundtrip with fourth-order Runge--Kutta nonlinear updates. An analytic pulse is relaxed for 2,000 noiseless roundtrips at $\delta=0.28$~rad, inside the single-soliton branch, and continued upwards in steps of 0.012194~rad to an imposed limit of 0.50~rad, where the soliton still exists. The detuning is then stepped backwards to $-0.30$~rad, mirroring the measured backward scan; Fig.~\ref{fig:soliton}a,b shows the interval covered by the measurement, 0.43 to $-0.22$~rad. Each step is followed by 400 roundtrips of evolution, increased to 1,000 below $\delta=0.21$~rad (Supplementary Note~3).

	For Fig.~\ref{fig:soliton}g, the converged solitons at four detunings are relaxed for another 2,000 roundtrips with noise and 1,500 without. Their carrier-frequency shift is removed before constructing perturbations and restored afterwards. The principal family scales both peak amplitude and inverse width, $B_s(\tau)=sB_*(s\tau)$ for a centred envelope, preserving $P_{\rm peak}T_0^2$ while varying the input energy. Each state is evolved for one physical roundtrip of the mean-field model without added noise. The energy change $\Delta E$ is the difference between output and input energies inside a pulse-centred gate extending $\pm2.5$ input-pulse FWHM, not the change integrated over the full numerical window. The higher zero crossing defines the stationary energy $E_*$ of this scalar response. Its local restoring fraction is $r=-\left.\partial\Delta E/\partial E\right|_{E_*}$, so a small energy perturbation evolves in this directional model as $\epsilon_{n+1}\simeq(1-r)\epsilon_n$ (Supplementary Note~6).
	
	To test the erbium contribution, the same perturbations are propagated with its gain frozen at the stationary value. The resulting restoring slope changes by approximately 2.5--2.9\%. For an isolated pulse with negligible background, $x=E_*/(T_RP_{\rm sat})\simeq(1.3$--$1.6)\times10^{-2}$, giving $E_*|dg/dE_*|=g_0x/(1+x)^2\simeq(6.1$--$7.4)\times10^{-3}$ per roundtrip, compared with the total restoring slope $r\simeq0.23$--0.29. This agrees with the frozen-gain comparison. 

	\paragraph{Interferometric phase acquisition.}
	For Fig.~\ref{fig:phase}, a tap of the 1550-nm master laser is combined with the oscillator output. Both fields propagate through 20~km of SMF for dispersive time stretching~\cite{Mahjoubfar2017}; their interference encodes the soliton spectral phase at the master frequency $\omega_p/2$, plus an uncalibrated differential-path offset. Beat frequencies approach 80~GHz near the 0.22-ns region surrounding the fringe waist; the chosen records were consequently sampled at 256~GS~s$^{-1}$, using the detection system described above. Separate $\pi$- and $2\pi$-resonance recordings each contain 64,537 consecutive roundtrips (approximately 7.5~ms) (Supplementary Note~7).

	\paragraph{Single-roundtrip phase extraction.}
	Every roundtrip is fitted independently over an unaveraged 0.6-ns segment centred on the fringe waist. The forward interferometric calculation underlying Fig.~\ref{fig:phase}f supplies three common templates: the stretched-pulse intensity $P(\tau)$ and two fixed orthogonal interference quadratures $U(\tau)$ and $V(\tau)$. The detected waveform is fitted to
	\[
	y(\tau)=g_{\rm fit}P(\tau)+c+aU(\tau)+bV(\tau),
	\qquad \psi_n=\arg(a+ib).
	\]
	Pulse width, accumulated dispersion and detector response are estimated from mean profiles and held fixed across roundtrips. A temporal offset is first selected by minimizing the least-squares residual over a 0.1-ps grid spanning $\pm20$~ps; the coefficients $g_{\rm fit}$, $c$, $a$ and $b$ are then fitted linearly. The shared dispersion is $-4.42\times10^{-22}$~s$^2$ and the Gaussian electrical amplitude response is $\exp[-(f/79.7~\mathrm{GHz})^2]$. Neither a repeating phase nor a $\pi$ parity step is imposed. This fit extracts one master-relative scalar phase per roundtrip. No trace was discarded by fitted fringe amplitude; the minimum observed amplitude was 0.83 of the model prediction. Fit diagnostics are given in Supplementary Note~7.
	
	\paragraph{Phase statistics and detection-noise floor.}
	For the $2\pi$ resonance the deterministic per-roundtrip advance is $\chi=0$; for the $\pi$ resonance it is $\chi=\pi$, corresponding to $A_{n+1}\simeq-A_n$. Phases are fitted before grouping by roundtrip parity. 
	
	For each lag $k$ in a record of $N=64{,}537$ roundtrips, Fig.~\ref{fig:phase}i plots the rms of $\psi_{n+k}-\psi_n-k\chi$ over the $N-k$ available pairs, after unwrapping $\psi_n-n\chi$. Every adjacent phase increment lies within $0.08\pi$ of its expected value, allowing unambiguous unwrapping in these records. Measured electrical noise propagated through the linearized phase estimator yields a per-roundtrip detection-noise rms $\sigma_{\rm det}=0.038$--0.040~rad. Assuming uncorrelated detection noise, the phase-difference floor is $D_0=\sqrt{2}\sigma_{\rm det}\simeq0.055$~rad. The observed total single-roundtrip phase rms is approximately 0.042~rad; it is distinct from the estimated detection-noise contribution and includes any real or path-related fluctuations. Noise correlations between fringe sub-bands are examined separately (Supplementary Note~8).
	For a Wiener process with Lorentzian relative linewidth $\Delta\nu$, the predicted lag dependence is $D_\psi^2(\tau)=D_0^2+2\pi\Delta\nu\tau$, with $\tau=kT_R$. At $\tau=3.5$~ms the measured $D_\psi\simeq0.061$~rad and $D_0\simeq0.055$~rad.

	\paragraph{Data availability.}
	The source data supporting the figures and the findings reported here will be deposited in Zenodo upon publication; the persistent identifier will be added when available.
	
	
	\paragraph{Code availability.}
	The simulation and analysis code will be released through GitHub upon publication; the repository URL will be provided when available. Solver configurations and fitted-parameter identifiers are listed in Supplementary Note~3.
	

	\paragraph{Acknowledgements.}  The authors thank Auro Perego for fruitful discussions.
	
	\paragraph{Funding.}
	This work was supported by Marie Sk\l{}odowska-Curie Actions (842676, 101149506, 101150387, 101202061); the European Research Council (101125625); an F.R.S.-FNRS Charg\'e de recherches fellowship (40010332); the F.R.S.-FNRS project EQP (40021523); and the FWO and F.R.S.-FNRS under the Excellence of Science programme (EOS, 40007560).
	\paragraph{Author contributions.}
	F.D.L. and F.L. conceived the project. F.D.L. led the experiments, with contributions from G.S., Y.S., C.D. and N.E. Y.S. led the numerical simulations, with contributions from G.S.; analysed the experimental data, with assistance from F.D.L.; synthesized the results and prepared the visualizations; and developed the narrative and wrote the original manuscript. F.L. and S.-P.G. supervised the project. All authors discussed the results and contributed to reviewing and editing the manuscript.
	\paragraph{Competing interests.}
	The authors declare no competing interests.
	\paragraph{Additional information.}
	Supplementary Information is available for this paper.
	
\end{document}